\documentclass[11pt]{article}
\usepackage [a4paper]{geometry}
\usepackage {ifxetex}
\ifxetex
\usepackage {fontspec}
\usepackage {xunicode}
\else
\usepackage [T1]{fontenc}
\usepackage {lmodern}
\usepackage [utf8]{inputenc}
\fi
\usepackage {amsmath,amssymb,amsfonts}
\usepackage {hyperref}
\usepackage {tikz}

\usepackage {dcolumn}
\newcolumntype {d}{D{.}{.}{-1}}

\usepackage{theorem}
\newtheorem {theorem}{Theorem}
\newtheorem {definition}[theorem]{Definition}
\newtheorem {defprop}[theorem]{Definition and Properties}
\newtheorem {proposition}[theorem]{Proposition}
\newtheorem {conjecture}[theorem]{Conjecture}
{
\theoremstyle {break}
\newtheorem {algorithm}[theorem]{Algorithm}
}
\newcommand {\inoutput}[2]{\textsc {Input:} #1\\
   \textsc {Output:} #2}
\newenvironment {remark}{\paragraph {Remark.}}{}

\makeatletter
\newenvironment{proof}[1][Proof]{\par
  \normalfont \topsep6pt plus6pt\relax
  \trivlist
  \item[\hskip\labelsep
        \scshape
    #1.]\ignorespaces
}{%
  \hspace*{\fill}$\square$\endtrivlist\@endpefalse
}

\def\fg{\ifdim\lastskip>\z@\unskip\fi~»}
\makeatother

\newcommand {\step}[1]{#1)}

\renewcommand {\theta}{\vartheta}
\renewcommand {\phi}{\varphi}
\renewcommand {\epsilon}{\varepsilon}
\renewcommand {\geq}{\geqslant}
\renewcommand {\leq}{\leqslant}
\newcommand{\lquo}[2]
{\leavevmode\kern-.1em\lower.25ex\hbox{$#2$}\kern-.1em\backslash\kern-.1em\raise.2ex\hbox{$#1$}}
\newcommand {\code}[1]{\texttt {#1}}
\newcommand {\parigp}{\textsc {Pari/Gp}}
\newcommand {\magma}{\textsc {Magma}}
\newcommand {\cmh}{\textsc {Cmh}}
\newcommand {\fplll}{\textsc {Fplll}}
\newcommand {\gmp}{\textsc {Gnu Mp}}
\newcommand {\mpfr}{\textsc {Gnu Mpfr}}
\newcommand {\mpfrcx}{\textsc {Mpfrcx}}
\newcommand {\mpc}{\textsc {Gnu Mpc}}
\newcommand {\Ot}{\ensuremath{\mathop{\tilde O}}}
\newcommand {\Kr}{K^r}
\newcommand {\Phir}{\Phi^r}
\newcommand {\phir}{\phi^r}

\newcommand {\Dr}{D^r}
\newcommand {\Ar}{A^r}
\newcommand {\Br}{B^r}

\newcommand {\Yr}{Y^r}
\newcommand {\Zr}{Z^r}

\newcommand {\yr}{y^r}
\newcommand {\zr}{z^r}
\newcommand {\F}{\mathbb {F}}
\newcommand {\Z}{\mathbb {Z}}
\newcommand {\Q}{\mathbb {Q}}
\newcommand {\R}{\mathbb {R}}
\ifxetex
\renewcommand {\C}{\mathbb {C}}
\else
\newcommand {\C}{\mathbb {C}}
\fi
\newcommand {\id}{\mathrm {id}}
\newcommand {\Cl}{\mathrm {Cl}}
\newcommand {\af}{\mathfrak {a}}
\newcommand {\bg}{\mathfrak {b}}
\newcommand {\pf}{\mathfrak {p}}
\newcommand {\qf}{\mathfrak {q}}
\newcommand {\Cf}{\mathfrak {C}}

\newcommand {\Fc}{\mathcal {F}}
\newcommand {\Hc}{\mathcal {H}}
\newcommand {\Mc}{\mathcal {M}}
\newcommand {\Oc}{\mathcal {O}}
\newcommand {\Qc}{\mathcal {Q}}

\newcommand {\Uc}{\mathcal {U}}
\newcommand {\Gal}{\operatorname {Gal}}
\newcommand {\Res}{\operatorname {Res}}
\newcommand {\Norm}{\operatorname {N}}
\newcommand {\kernel}{\operatorname {ker}}
\newcommand {\image}{\operatorname {im}}
\newcommand {\Tr}{\operatorname {Tr}}
\newcommand {\Sp}{\operatorname {Sp}}
\newcommand {\mult}{\operatorname {M}}
\newcommand {\Hhat}{\hat H}

\title {Computing class polynomials for abelian surfaces}
\author {Andreas Enge\footnote {INRIA, LFANT, F-33400 Talence, France \newline
CNRS, IMB, UMR 5251, F-33400 Talence, France \newline
Univ. Bordeaux, IMB, UMR 5251, F-33400 Talence, France \newline
andreas.enge@inria.fr}
\ and
Emmanuel Thomé\footnote {INRIA, CARAMEL, Nancy, France \newline
emmanuel.thome@inria.fr}}
\date {9 December 2013}

\begin {document}
\maketitle

\begin {abstract}
We describe a quasi-linear algorithm for computing Igusa class polynomials
of Jacobians of genus~$2$ curves via complex floating-point approximations
of their roots. After providing an explicit treatment of the computations
in quartic CM fields and their Galois closures, we pursue an approach due
to Dupont for evaluating $\theta$-constants in quasi-linear time using
Newton iterations on the Borchardt mean. We report on experiments with
our implementation and present an example with class number~$20016$.
\end {abstract}

\section {Introduction}

Igusa class polynomials describe the complex multiplication points in the
moduli space of principally polarised abelian surfaces, that is, they
parameterise abelian varieties of dimension~$2$ with complex multiplication
by a maximal order of a quartic CM field. Such abelian surfaces are
Jacobians of hyperelliptic curves of genus~$2$, so that by computing Igusa
class polynomials one may obtain genus~$2$ curves over finite fields
with known Jacobian cardinality. 

In the dimension~$1$ case of elliptic curves, several approaches have been
described in the literature. While the output of the algorithms (a large
polynomial) is of exponential size in the input (a number field described
by a single integer), all of these approaches may lead to an algorithm
with a complexity that is quasi-linear (up to logarithmic factors) in its
output size: The complex analytic method uses floating point approximations
to the roots of the class polynomials \cite {Enge09}; the $p$-adic
approach starts from a curve with the given endomorphism ring over a small
finite field and lifts its invariants to a $p$-adic field
\cite {CoHe02,Broker08}; the Chinese remaindering approach combines curves
over several small prime fields \cite {BeBrEnLa08}.

In principle, the same approaches apply to abelian surfaces. A $2$-adic
algorithm is described in \cite{GaHoKoRiWe06}, and there are currently
attempts at making the Chinese remainder based method more
efficient~\cite{LaRo13}.
So far, $2$-adic lifting appears to have been the most successful approach:
The Echidna database maintained by
Kohel\footnote{\url{http://echidna.maths.usyd.edu.au/echidna/dbs/complex_multiplication2.html}}
contains Igusa class polynomials, the largest of which is of degree~$576$
and has been obtained by lifting from a curve
over $\F_{2^6}$\footnote {Personal communication}.

A detailed description of the complex analytic approach, together with
complexity analyses
of its different steps, has recently been given in~\cite{Streng09,Streng10}.
Our work pushes the limits for the
attainable degrees of Igusa class polynomials: We present an example of
degree $20\,016$. Moreover, relatively small class
polynomials (say, below degree $150$) can be computed in matters of
seconds. The key tool in this approach is the use of a quasi-linear
algorithm for the computation of $\theta$-constants, initially described
in~\cite{Dupont06}.

This article is organised as follows. \S\ref{sec:cm} presents the
necessary background material for discussing the complex
multiplication theory of abelian surfaces and states the general algorithm.
\S\S\ref{sec:explicit} and~\ref{sec:shimuragroup} show how to explicitly
(providing concrete descriptions for the occurring number fields, maps
between them and their embeddings) and symbolically compute an appropriate
set of reduced period matrices,
which form the input of the computationally expensive step of computing
$\theta$-constants, detailed in \S\ref{sec:computing-thetas}. The
recognition as algebraic numbers of the coefficients of the Igusa class
polynomials from their
approximations by complex embeddings is described in \S\ref{sec:lll}, and
experimental results are given in \S\S\ref{sec:implementation}
and~\ref{sec:large}.

All computations presented in this article have been
achieved with the software package~\cmh\cite {cmh},
released under the GNU General Public License.

\section {Complex multiplication theory}
\label {sec:cm}

In this section, we provide a concise introduction to the theory of complex
multiplication of principally polarised abelian surfaces or, equivalently,
Jacobians of genus~$2$ hyperelliptic curves over the complex numbers, to
the extent needed to describe our algorithms and implementation.
The presentation follows \cite{Streng10},
and proofs are given in \cite{ShTa61,Shimura98,Streng10,Streng09}.

\subsection {Quartic CM fields and abelian surfaces}
\label {ssec:cmfields}

A \emph {CM field} $K$ is an imaginary-quadratic extension of a
totally real number field $K_0$. We denote by $\kappa$ indiscriminately
the complex conjugation on~$\C$ and the automorphism generating
$\Gal (K / K_0)$. For any embedding $\phi : K \to \C$, we have
$\kappa \circ \phi = \phi \circ \kappa$, which justifies the notation
$\overline \phi = \kappa \circ \phi$.

\emph {Quartic CM fields} $K$ of degree~$4$ over~$\Q$ come in three Galois
types. Generically, $K / \Q$ is not Galois, the Galois closure $L / K$ is
of degree~$2$, and $\Gal (L / \Q)$ is isomorphic to the dihedral group~$D_4$.
The Galois closure $L$ is itself a CM field, and the complex conjugation of~$L$, which we
denote again by~$\kappa$, restricts to the complex conjugation of~$K$.
If $K / \Q$ is Galois, it may be either cyclic or biquadratic. We will not
consider the biquadratic case in the following, since then the abelian
surfaces of which it is the endomorphism algebra are products of elliptic
curves; so from now on, all Galois quartic CM fields are tacitly understood
to be cyclic.

A \emph {CM type} of a quartic CM field~$K$ is a set
$\Phi = \{ \phi_1, \phi_2 \}$ of two embeddings $K \to \C$ such that
$\phi_2 \neq \overline \phi_1$; that is, it contains one out of each pair
of complex-conjugate embeddings. Two CM types $\Phi$ and $\Phi'$ are
equivalent if there is an automorphism $\sigma$ of~$K$ such that
$\Phi' = \Phi \circ \sigma$; in particular, $\Phi$ and $\overline \Phi$ are
equivalent. If $K / \Q$ is Galois, there is only one equivalence class
of CM types; otherwise, there are two inequivalent classes
$\Phi = \{ \phi_1, \phi_2 \}$ and $\Phi' = \{ \phi_1, \overline \phi_2 \}$.

For a given CM type $\Phi = \{ \phi_1, \phi_2 \}$, its
\emph {reflex field} is the field $\Kr$ generated over~$\Q$ by the
\emph {type traces}, that is,
$\Kr = \Q \big( \{ \phi_1 (x) + \phi_2 (x) : x \in K \} \big)$;
it is itself a quartic CM field and we denote by $\Kr_0$ its
real-quadratic subfield. Equivalent CM types yield conjugate reflex
fields. In the Galois case, $K$ and $\Kr$ are isomorphic, while in the
dihedral case, they are not isomorphic, but the two reflex fields for
the two inequivalent CM types are.
In both cases, there is a natural way of defining a dual CM type
$\Phir = \{ \phir_1, \phir_2 \}$ of~$\Kr$, and the reflex field of $\Kr$
is isomorphic to~$K$.
Define the (dual) \emph {type norm}
$\Norm_{\Phir} : \Kr \to K$ by
$x \mapsto \phir_1 (x) \phir_2 (x)$, so that
\begin {equation}
\label {eq:typenormnorm}
\Norm_{\Phir} \overline \Norm_{\Phir} = \Norm;
\end {equation}
this map extends to ideals and ideal classes.

In \S\ref {sec:explicit}, we provide explicit equations for all
occurring number fields and consider their embeddings from an effective
point of view.

Let $\af$ be a fractional ideal of $\Oc_K$. A CM type
$\Phi = \{ \phi_1, \phi_2 \}$ induces an embedding
$K \to \C^2$, $x \mapsto (\phi_1 (x), \phi_2 (x))$, under which
$\Phi (\af)$ is a lattice of rank~$4$. Its cokernel $\C^2 / \Phi (\af)$,
a complex torus of genus~$2$, is an \emph {abelian surface}.
Let $\delta_K^{-1} = \{ y \in K : \Tr (x y) \in \Z \:\: \forall x \in \Oc_K \}$
be the codifferent ideal of~$K$. Assume that $(\af \overline \af \delta_K)^{-1}$
is principal and generated by some $\xi \in K$ such that
$\phi_1 (\xi), \phi_2 (\xi) \in i \R^{> 0}$;
in particular, $\xi \overline \xi \in K_0$ is totally negative. Then
$E_{\Phi, \xi} : \Phi (K)^2 \to \Q,
(\Phi (x), \Phi (y)) \mapsto \Tr (\xi \overline x y)$
is a symplectic form over~$\Q$ which takes integral values on $\Phi (\af)^2$.
By tensoring with~$\R$, one obtains a symplectic form $\C^2 \to \R$
such that $(x, y) \mapsto E_{\Phi, \xi} (i x, y)$ is symmetric and
positive definite, a \emph {principal polarisation} on
$\C^2 / \Phi (\af)$.

The principally polarised abelian surface
$A (\Phi, \af, \xi) = \left( \C^2 / \Phi (\af), E_{\Phi, \xi} \right)$
has complex multiplication by~$\Oc_K$; conversely, any such surface can be
obtained up to isomorphism in this way. Two principally polarised abelian
surfaces $A (\Phi, \af, \xi)$ and $A (\Phi', \af', \xi')$ are isomorphic
if and only if $\Phi = \Phi'$ (up to equivalence) and there is a
$u \in K^\ast$ such that $\af' = u \af$ and $\xi' = (u \overline u)^{-1} \xi$.
In particular this implies that $u \overline u \in K_0$ is totally positive,
and that we may assume $\af$ to be an integral ideal of $\Oc_K$.

\subsection {The Shimura group, its type norm subgroup and cosets}
\label {ssec:shimuragroup}

The Igusa invariants to be defined in \S\ref {ssec:igusa} determine the
moduli space $\Mc$ of principally polarised complex abelian surfaces,
which has a model over~$\Q$. Let $\Mc_{K, \Phi}$ be the subset of
surfaces $A (\Phi, \af, \xi)$ obtained from an integral ideal of $\Oc_K$
and the CM type $\Phi$ as described in \S\ref {ssec:cmfields}.
Then $\Mc_{K, \Phi}$ is stable under $\Gal (\overline \Q / \Kr_0)$.
If $K$ is cyclic, then $\Mc_{K, \Phi}$ is even stable under
$\Gal (\overline \Q / \Q)$. Otherwise let $\Phi'$ be inequivalent with~$\Phi$.
Then $\Mc_{K, \Phi}$ and $\Mc_{K, \Phi'}$ are disjoint and conjugate under
$\Gal (\Kr_0 / \Q)$ \cite[Lemmata~1.1 and~2.1]{Streng10}.

Let the \emph {Shimura class group} $\Cf$ be defined by
\begin {equation}
\label {eq:shimuragroup}
\Cf = \big\{ (\af, u) :
\af \text { a fractional ideal of } \Oc_K,
\af \overline \af = u \Oc_K, \text { and }
u \in K_0
\text { totally positive} \} / \sim
\end {equation}
with component-wise multiplication. The equivalence relation denoted
$\sim$ above is the one
induced by principal ideals, more precisely the equivalence modulo the subgroup given by the
$(v \Oc_K, v \overline v)$ with $v \in K^\ast$ and $v \overline v \in K_0$
totally positive.

By the discussion of \S\ref {ssec:cmfields}, the Shimura class group
$\Cf$ acts regularly
on $\Mc_{K, \Phi}$ via
\begin {equation}
\label {eq:shimura}
(\bg, u) \cdot A (\Phi, \af, \xi) = A (\Phi, \bg^{-1} \af, u \xi).
\end {equation}
Consider the dual type norm map
$\Norm_{\Phir} : \Cl_{\Kr} \to \Cf,
\bg \mapsto \left( \Norm_{\Phir} (\bg), \Norm (\bg) \right),$
which is well defined by \eqref {eq:typenormnorm}.
For any $A (\Phi, \af, \xi)$, the action induced by
$\Norm_{\Phir} (\Cl_{\Kr})$ is that of the Galois group of the field of
moduli of $A (\Phi, \af, \xi)$ over~$\Kr$
\cite[Theorem~9.1]{Streng10};
otherwise said, the field of moduli is the fixed field of
$\kernel (\Norm_{\Phir})$ inside the Hilbert class field of~$\Kr$.
The cokernel of $\Norm_{\Phir}$ is elementary abelian of exponent~$1$
or~$2$ \cite[Theorem~2.2]{Streng10}, so $\Mc_{K, \Phi}$ splits into orbits
under~$\Cf$ of size $|\image (\Norm_{\Phir})|$, and the number of orbits
is a power of~$2$. As stated above, these orbits are in fact defined
over~$\Kr_0$, with the orbits of $\Mc_{K, \Phi}$ and $\Mc_{K, \Phi'}$ being
mapped to each other by $\Gal (\Kr_0 / \Q)$.

\subsection {\texorpdfstring {$\theta$}{Theta}-functions, Igusa invariants
and class polynomials}
\label {ssec:igusa}

Given an ideal $\af$ and a principal polarisation $E_{\Phi, \xi}$ as in
\S\ref {ssec:cmfields}, one may choose a $\Z$-basis
$(\alpha_1, \alpha_2, \alpha_3, \alpha_4)$ of~$\af$ such that
$v_1 = \Phi (\alpha_1)$, $v_2 = \Phi (\alpha_2)$, $w_1 = \Phi (\alpha_3)$,
$w_2 = \Phi (\alpha_4)$ form a symplectic basis, for which
$E_{\Phi, \xi}$ becomes
$\begin {pmatrix} 0 & \id_2 \\ -\id_2 & 0 \end {pmatrix}$.
That the change of basis is defined over $\Z$ and not only over $\R$ follows
from the principality of the polarisation; we also call this basis of~$\af$
\textit {symplectic}.
Let $V = \begin {pmatrix} v_1 & v_2 \end {pmatrix}$,
$W = \begin {pmatrix} w_1 & w_2 \end {pmatrix} \in \C^{2 \times 2}$.
Rewriting the ambient vector space $\C^2$ and $\Phi (\af)$ in the basis
spanned by $w_1$ and $w_2$, we obtain
$\Phi (\af)
= \begin {pmatrix} \Omega_{\Phi, \af, \xi} & \id_2 \end {pmatrix} \Z^4$
with the \emph {period matrix}
\begin {equation}
\label {eq:periodmatrix}
\Omega_{\Phi, \af, \xi} = W^{-1} V
\end {equation}
in the \emph {Siegel half space}
$\Hc_2 = \left\{ \Omega \in \C^{2 \times 2} :
\Omega \text { symmetric and } \Im (\Omega) \text { positive definite} \right\}$.
The symplectic group $\Sp_4 (\Z)$ acts on $\Hc_2$ by
\[
\begin {pmatrix} A & B \\ C & D \end {pmatrix} \Omega
= (A \Omega + B) (C \Omega + D)^{-1},
\]
where $A$, $B$, $C$, $D \in \Z^{2 \times 2}$. As in the case of genus~$1$,
a fundamental domain for $\Hc_2$ exists under the action of $\Sp_4(\Z)$.
Reduction into the fundamental domain is
discussed in~\S\ref{ssec:symbolic-reduction}.

The $\theta$-constants are certain modular forms of weight $1/2$ for
$\Sp_4 (\Z)$.
Let $a = \begin {pmatrix} a_1 \\ a_2 \end {pmatrix}$,
$b = \begin {pmatrix} b_1 \\ b_2 \end {pmatrix}
\in \left( \frac {1}{2} \Z \right)^2$ be two vectors of
\emph{$\theta$-characteristics}.
Then for $\Omega \in \Hc_2$,
\begin{equation}
\label{eq:theta}
\theta_{16 a_1 + 8 a_2 + 4 b_1 + 2 b_2} (\Omega)
= \theta_{a, b} (\Omega)
= \sum_{n \in \Z^2} e^{2 \pi i
     \left( \frac {1}{2} (n + a)^\intercal \Omega (n + a)
        + (n + a)^\intercal b \right)}.
\end{equation}
Only the \emph {even} $\theta$-constants $\theta_i$ for
$i \in T = \{ 0, 1, 2, 3, 4, 6, 8, 9, 12, 15 \}$ are not identically~$0$.

The following \emph {duplication formulæ} relate the values of the
squares of the ten even $\theta$-constants in the argument~$\Omega$ with
the values of the four \emph {fundamental $\theta$-constants}
$\theta_0, \ldots, \theta_3$ (which have $a = 0$) in the argument
$\Omega / 2$ (omitted from the formulæ for the sake of conciseness).
\begin {equation}
\label {eq:duplication}
\begin{split}
    4\theta_0^2(\Omega) &= \theta_0^2 + \theta_1^2 + \theta_2^2 + \theta_3^2\\
    4\theta_1^2(\Omega) &= 2\theta_0\theta_1 + 2\theta_2\theta_3\\
    4\theta_2^2(\Omega) &= 2\theta_0\theta_2 + 2\theta_1\theta_3\\
    4\theta_3^2(\Omega) &= 2\theta_0\theta_3 + 2\theta_1\theta_2\\
    4\theta_4^2(\Omega) &= \theta_0^2 - \theta_1^2 + \theta_2^2 - \theta_3^2
\end{split}
\qquad
\begin{split}
    4\theta_6^2(\Omega) &= 2\theta_0\theta_2 - 2\theta_1\theta_3\\
    4\theta_8^2(\Omega) &= \theta_0^2 + \theta_1^2 - \theta_2^2 - \theta_3^2\\
    4\theta_9^2(\Omega) &= 2\theta_0\theta_1 - 2\theta_2\theta_3\\
 4\theta_{12}^2(\Omega) &= \theta_0^2 - \theta_1^2 - \theta_2^2 + \theta_3^2\\
 4\theta_{15}^2(\Omega) &=  2\theta_0\theta_3 - 2\theta_1\theta_2
\end{split}
\end {equation}
Denote by $h_j$ the following modular forms of weight~$j$:
\begin {equation}
\label {eq:h}
\begin {split}
& h_4 = \sum_{i \in T} \theta_i^8, \quad
   h_6 = \sum_{60 \text { triples } (i, j, k) \in T^3}
         \pm (\theta_i \theta_j \theta_k)^4, \\
& h_{10} = \prod_{i \in T} \theta_i^2, \quad
   h_{12} = \sum_{15 \text { tuples } (i, j, k, l, m, n) \in T^6}
            (\theta_i \theta_j \theta_k \theta_l \theta_m \theta_n)^4;
\end {split}
\end {equation}
for the exact definitions, see \cite[{\S}II.7.1]{Streng10}.
These generate the ring of holomorphic Siegel modular forms over~$\C$, see
\cite[Corollary p. 195]{Igusa62} and \cite[Remark~7.2]{Streng10}.
The moduli space of principally polarised abelian surfaces is of
dimension~$3$ and parameterised by \emph {absolute Igusa invariants},
modular functions (thus of weight~$0$) in
$\Z \left[ h_4, h_6, h_{12}, h_{10}^{-1} \right]$.
Different sets of invariants have been suggested in the literature.
The most cited one is Spallek's, who uses a system in the linear span of
$\frac {h_{12}^5}{h_{10}^6}$, $\frac {h_{12}^3 h_4}{h_{10}^4}$,
$\frac {h_{12}^2 h_6}{h_{10}^3}$
\cite[Satz~5.2]{Spallek94}.
Streng defines invariants with the minimal powers of $h_{10}$ in the
denominator as
\begin {equation}
\label {eq:igusa}
j_1 = \frac {h_4 h_6}{h_{10}}, \quad
j_2 = \frac {h_4^2 h_{12}}{h_{10}^2}, \quad
j_3 = \frac {h_4^5}{h_{10}^2}.
\end {equation}

The principally polarised abelian surfaces $A (\Phi, \af, \xi)$ are
parameterised by the triples of \emph {singular values}
$\left( j_1 (\Omega), j_2 (\Omega),
j_3 (\Omega) \right)$ in the period matrices
$\Omega=\Omega_{\Phi, \af, \xi}$, which may be obtained from the action of the
Shimura class group $\Cf$ on a fixed \emph {base point}
$\beta = (\Phi, \af_\Phi, \xi_\Phi)$.
The singular values lie in the subfield of the Hilbert class field
of $\Kr$ given in \S\ref {ssec:shimuragroup}. Following the discussion
there, the \emph {Igusa class polynomials}
$I_i (X) = \prod_{(\Phi, \af, \xi)}
\left( X - j_i (\Omega_{\Phi, \af, \xi}) \right)$
are defined over~$\Q$. More precisely their irreducible factors, over~$\Kr_0$
in the dihedral case or~$\Q$ in the cyclic case, are given by
$$\prod_{C \in \Norm_{\Phir} (\Cl_{\Kr})}
   \left( X - j_i (\Omega_{C C' \cdot \beta}) \right)\text,$$
where $\Phi$ is one CM type and $C' \in \Cf / \Norm_{\Phir} (\Cl_{\Kr})$.

In the following, we fix a CM type $\Phi$ (for its explicit description,
see \S\ref {sec:explicit}) and a base point
$\beta=(\Phi, \af_\Phi, \xi_\Phi)$ and let
\begin {equation}
\label {eq:H}
H_1 (X) = \prod_{C \in \Norm_{\Phir} (\Cl_{\Kr})}
\left( X - j_1 (\Omega_{C \cdot \beta}) \right).
\end {equation}
As elements of the same class field, the singular values of~$j_2$
and~$j_3$ are rational expressions in the singular value of~$j_1$.
Computationally, it is preferable to use the \emph {Hecke representation}
in the trace-dual basis to keep denominators small. We thus define
polynomials $\Hhat_{2}$ and $\Hhat_3$ through
$j_i H_1' (j_1) = \Hhat_i (j_1)$ with
\begin {equation}
\label {eq:Hhat}
\Hhat_i (X) = \sum_{C \in \Norm_{\Phir} (\Cl_{\Kr})} j_i
   (\Omega_{C \cdot \beta})
\prod_{D \in \Norm_{\Phir} (\Cl_{\Kr}) \backslash \{ C \}}
   \left( X - j_1 (\Omega_{D\cdot \beta}) \right)
\end {equation}
for $i \in \{ 2, 3 \}$, where
$H_1$, $\Hhat_2$, $\Hhat_3 \in \Kr_0 [X]$ in the dihedral case and
$\in \Q [X]$ in the cyclic case.

\subsection {Algorithm for Igusa class polynomials}

We briefly summarise the algorithm for computing class polynomials.

\begin {algorithm}
\label{algo:igusa-classpol}
\inoutput {CM field $K$ and CM type $\Phi = \{ \phi_1, \phi_2 \}$ of $K$}
 {Irreducible class polynomials $H_1$, $\Hhat_2$, $\Hhat_3 \in \Kr_0 [X]$
  in the dihedral case and $\in \Q [X]$ in the Galois case}
\begin {enumerate}
\item
Compute $\Norm_{\Phir} (\Cl_{\Kr})
   = \{ (\bg_1, u_1), \ldots, (\bg_h, u_h) \} \subseteq \Cf$.
\item
Compute a base point $\beta=(\Phi, \af_\Phi, \xi_\Phi)$ such that
$\left\{
    \begin{array}{l}
   (\af_\Phi \overline \af_\Phi \delta_K)^{-1} = (\xi_\Phi),\\
   \phi_1 (\xi_\Phi), \phi_2 (\xi_\Phi) \in i \R^{>0}.
    \end{array}\right.$
\item
    Enumerate $\{C\cdot \beta=(\Phi, \bg_i^{-1} \af_\Phi, u_i
    \xi_\Phi) : \ C=(\bg_i,u_i)\in\Norm_{\Phir} (\Cl_{\Kr})\}$ and compute
   the associated period matrices
   $\Omega_i = \Omega_{C\cdot\beta}$
   for $i = 1, \ldots, h$.
\item
For $i = 1, \ldots, h$, compute the fundamental $\theta$-constants
$\theta_0 (\Omega_i / 2), \ldots, \theta_3 (\Omega_i / 2)$;
then deduce the squares of the ten even $\theta$-constants
$\theta_k^2 (\Omega_i)$ by~\eqref {eq:duplication},
the values $h_k (\Omega_i)$ by~\eqref {eq:h} and finally
the triples
   $J_i = \big( j_1 (\Omega_i), j_2 (\Omega_i), j_3 (\Omega_i) \big)$
by~\eqref {eq:igusa}.
\item
Let $H_1 = \prod_{i=1}^h (X - J_{i, 1})$,
   $\Hhat_k = \sum_{i=1}^h J_{i,k} \prod_{l \neq i} (X - J_{l,1})
   \in \C [X]$ for $k \in \{ 2, 3 \}$.
\item
Recognise the coefficients of $H_1$, $\Hhat_2$, $\Hhat_3$ as elements
   of $\Kr_0$ or $\Q$, respectively.
\end {enumerate}
\end {algorithm}

The different steps of the algorithm and our implementation are detailed in
the following chapters. The symbolic computations related to number fields in
Steps~\step {1} and~\step {2} and to the period matrices $\Omega_i$
in Step~\step {3} are described in \S\ref {sec:explicit}.
Step~\step {1} is treated in \S\ref {sec:typenorm},
Step~\step {4} in \S\ref {sec:computing-thetas} and
Step~\step {6} in \S\ref{sec:lll}.

\section {Explicit equations and symbolic period matrices}
\label {sec:explicit}

While Algorithm \ref{algo:igusa-classpol} \textit {in fine} works with
complex approximations obtained \textit {via} CM types, it starts from an
algebraic setting. In this section, we examine how to carry out the
computations as far as possible symbolically with algebraic numbers,
which relieves us from the need to decide on the necessary precision early on.
In particular, in \S\ref {ssec:period} we replace the complex embeddings
forming a CM type by algebraic embeddings into the compositum~$L$ of all
involved fields, followed by a ``universal'' embedding $\psi$ of~$L$
into~$\C$. Taking preimages under~$\psi$, the entries of the period
matrices $\Omega \in \C^{2 \times 2}$ may then be interpreted as elements
of the reflex field and may be handled symbolically.
We then fix a model for the CM field~$K$ in~\S\ref {ssec:nf} and
derive explicit equations for all considered fields and embeddings.

Recall the notation of~\S\ref {sec:cm}:
$K$ is a quartic CM field, $K_0$ its real quadratic subfield and $L$
its Galois closure with Galois group $G$. We consider only the dihedral case
$[L : K] = 2$ and $G = D_4$ and the cyclic case $L = K$ and $G = C_4$.
Let $\Phi = (\phi_1, \phi_2)$ be a CM type, where
$\phi_1$, $\phi_2 : K \to \C$ are two complex embeddings of~$K$ with
$\phi_2 \neq \overline \phi_1$, and let $\Kr$ be the reflex field
of $K$ with respect to~$\Phi$.

\subsection {Galois theory, embeddings and period matrices}
\label {ssec:period}

\subsubsection {The dihedral case}
\label {sssec:perioddihedral}

\paragraph {Galois theory.}

Let $K=\Q(y)$ be a non-Galois quartic CM field. The following statements
are easily seen to be true when choosing a generator $y$ such that $z = y^2$
belongs to the real subfield $K_0$, so that
$K = \Q [Y] / (Y^4+AY^2+B)$ for some $A, B \in \Q$.
The Galois closure of $K$ is then $L=K(y')=\Q(y,y')$, where the roots of
the minimal polynomial of~$y$ in $L$ are $\pm y$ and $\pm y'$
(the former could be identified with $\phi_1 (y)$ in \eqref {eq:phiy},
the latter with $\phi_2 (y)$).
The automorphisms in $G = \Gal (L / \Q)$ are uniquely determined by their
images on $y$ and $y'$, and we obtain the following diagram of fields
and Galois groups:
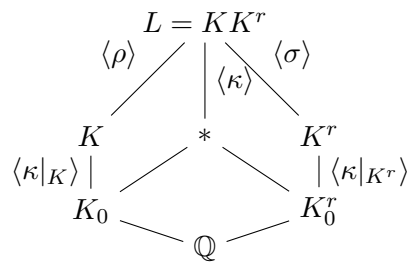
\begin{center}
\begin {tikzpicture}
\node (Q)   at (2, 0.5) {$\Q$};
\node (K0)  at (0.5, 1) {$K_0$};
\node (K)   at (0.5, 2) {$K$};
\node (K0r) at (3.5, 1) {$\Kr_0$};
\node (Kr)  at (3.5, 2) {$\Kr$};
\node (k)   at (2, 2) {$\ast$};
\node (L)   at (2, 3.5) {$L = K \Kr$};
\draw (Q) -- (K0) -- node [left]{$\langle \kappa|_K \rangle$} (K)
                  -- node [above left]{$\langle \rho \rangle$} (L);
\draw (Q) -- (K0r) -- node [right]{$\langle \kappa|_{\Kr} \rangle$} (Kr)
                   -- node [above right]{$\langle \sigma \rangle$} (L);
\draw (K0) -- (k) -- node [right]{$\langle \kappa \rangle$} (L);
\draw (K0r) -- (k);
\end {tikzpicture}
\end{center}
Here the automorphisms are given by
\begin {align*}
\rho   & : y \mapsto y, \; y' \mapsto -y'
\text { of order } 2 \\
\sigma & : y \mapsto y', \; y' \mapsto y
\text { of order $2$, which fixes the generator $y + y'$ of $\Kr$} \\
\tau   & : y \mapsto y', \; y' \mapsto -y
\text { of order $4$} \\
\kappa & = \tau^2 : y \mapsto -y, \; y' \mapsto -y
\text { is the complex conjugation}.
\end {align*}
So $G$ is the dihedral group $D_4$ with generators $\tau$ of order~$4$
and $\rho$ (or $\sigma$) of order~$2$ and additional relation
$\rho \tau \rho = \tau^3$,
and with $\langle \kappa \rangle$ as its centre.

\paragraph {Embeddings and CM types.}
There is a unique embedding $\psi : L \to \C$ such that
$\phi_1 = \psi|_K$ and $\phi_2 = (\psi \sigma)|_K$
(where multiplication denotes composition), which can be seen as follows.
First of all, there are two embeddings which, restricted to~$K$,
yield $\phi_1$; we denote them by $\psi_1$ and $\psi_1' = \psi_1 \rho$.
Now there is $s \in G$, uniquely defined up to multiplication
by $\rho$ from the right, such that $\phi_2 = (\psi_1 s)|_K$.
Since $\phi_2 \neq \phi_1$ and $\phi_2 \neq \overline \phi_1$, the
automorphism~$s$ is neither $1$, $\rho$, $\kappa = \tau^2$ nor
$\kappa \rho = \tau^2 \rho$.
This leaves $s$ as one of $\tau = \rho \sigma$,
$\tau \rho = \rho \sigma \rho$, $\tau^3 = \sigma \rho$ or
$\tau^3 \rho = \sigma$.
If $s|_K = \sigma|_K = (\sigma \rho)|_K$, we may choose $\psi = \psi_1$.
Otherwise, $s|_K = \rho \sigma$, and
$(\psi_1' \sigma)|_K = (\psi_1 \rho \sigma)|_K = (\psi_1 s)|_K = \phi_2$,
so we choose $\psi = \psi_1'$.

\paragraph {Period matrices.}
Let $(\alpha_1, \ldots, \alpha_4)$ be a symplectic basis for the
ideal~$\af$ of~$K$ with respect to $E_{\Phi, \xi}$ as defined
in~\S\ref {ssec:igusa}. Then

\begin {eqnarray*}
V & = &
\begin {pmatrix}
\phi_1 (\alpha_1) & \phi_1 (\alpha_2) \\
\phi_2 (\alpha_1) & \phi_2 (\alpha_2)
\end {pmatrix}
= \psi \left(
\begin {pmatrix}
\alpha_1        & \alpha_2 \\
\alpha_1^\sigma & \alpha_2^\sigma
\end {pmatrix}
\right), \\
W & = &
\begin {pmatrix}
\phi_1 (\alpha_3) & \phi_1 (\alpha_4) \\
\phi_2 (\alpha_3) & \phi_2 (\alpha_4)
\end {pmatrix}
= \psi \left(
\begin {pmatrix}
\alpha_3        & \alpha_4 \\
\alpha_3^\sigma & \alpha_4^\sigma
\end {pmatrix}
\right)
\end {eqnarray*}
and
\begin {equation}
\label {eq:period}
\Omega_{\Phi, \af, \xi} = W^{-1} V
= \psi (M)
\text { with }
M =
\frac {1}{\alpha_3 \alpha_4^\sigma - \alpha_4 \alpha_3^\sigma}
\begin {pmatrix}
\alpha_4 \alpha_1^\sigma - \alpha_1 \alpha_4^\sigma &
\alpha_4 \alpha_2^\sigma - \alpha_2 \alpha_4^\sigma \\
\alpha_3 \alpha_1^\sigma - \alpha_1 \alpha_3^\sigma &
\alpha_3 \alpha_2^\sigma - \alpha_3 \alpha_2^\sigma
\end {pmatrix}
\end {equation}
by \eqref {eq:periodmatrix}.
The entries of $M$ are invariant under $\sigma$ and thus elements of~$\Kr$.

\begin {remark}
It is crucial to choose out of the two embeddings $\psi : L \to \C$
that extend $\phi_1$ the one compatible with $\phi_2$.
The other one corresponds to the second CM type
$\Phi' = (\phi_1, \overline \phi_2)$ with reflex field $(\Kr)'$ and
$\Gal \left( L / (\Kr)' \right) = \langle \kappa \sigma \rangle
= \langle \rho \sigma \rho \rangle$.
\end {remark}

\subsubsection {The cyclic case}
Here we have the much simpler situation
\begin {center}
\begin {tikzpicture}
\node (Q)   at (0, 0) {$\Q$};
\node (K0)  at (0, 1) {$K_0$};
\node (K)   at (0, 2) {$K$};
\draw (Q) -- (K0) --
   node [right]{$\langle \kappa \rangle = \langle \sigma^2 \rangle$} (K);
\end {tikzpicture}
\end {center}
We may choose $\psi = \phi_1$. Then there is a uniquely determined
$\sigma \in \Gal (K/\Q)$
such that $\phi_2 = \phi_1 \sigma$, and trivially $M$ of \eqref {eq:period}
has entries in~$\Kr$. In general, they will not lie in a subfield:
Since $\sigma$ is neither the identity nor complex conjugation, it is of
order~$4$.

\subsection {Number field computations}
\label {ssec:nf}

In this section we show how to express the elements of the reflex field
$\Kr$ and the normal closure $L$ in consistent ways, so as to be able
to compute type norms and entries of period matrices as given
by~\eqref{eq:periodmatrix}.
We use the same notation for elements of the Galois group $G$ of $L / \Q$
as in \S\ref {ssec:period}.

\subsubsection {The dihedral case}

\paragraph {Field equations.}
By choosing generating elements as in \S\ref {sssec:perioddihedral}
we may assume that
\begin {align*}
K_0   = \Q (z) &= \Q [Z] / \left( Z^2 + A Z + B \right) \text { with }
   A, B \in \Z^{> 0}, A^2 - 4 B > 0; \\
K     = \Q (y) &= \Q [Y] / \left( Y^4 + A Y^2 + B \right).
\end {align*}
We then select the CM type $\Phi = (\phi_1, \phi_2)$ with
\begin {equation}
\label {eq:phiy}
\phi_1 (y) = i \sqrt {\frac {A + \sqrt {A^2 - 4 B}}{2}}, \quad
\phi_2 (y) = i \sqrt {\frac {A - \sqrt {A^2 - 4 B}}{2}},
\end {equation}
where all the real roots are taken to be positive; the other CM type
is $\Phi' = (\phi_1, \overline \phi_2)$ with
$\overline \phi_2 (y) = - \phi_2 (y)$.
Recall from \S\ref {sssec:perioddihedral} the notations
$\Gal (L/K) = \langle \rho \rangle$,
$\Gal (L/\Kr) = \langle \sigma \rangle$,
and  let $\psi : L \to \C$ be such that
$\phi_1 = \psi|_K$ and $\phi_2 = (\psi \sigma)|_K$.
The reflex field $\Kr$ is generated by the type traces of~$K$;
letting $\yr=y+y^\sigma$, the equality
\begin {equation}
\label {eq:psiyp}
\psi (\yr) = \psi (y) + (\psi \sigma) (y) = \phi_1 (y) + \phi_2 (y)
\end {equation}
shows that we may consider $\yr$ as a generator of~$\Kr$.
This gives the equations
\begin {align*}
\Kr_0 = \Q (\zr) & =  \Q [\Zr] / \left( (\Zr)^2 + \Ar \Zr + \Br \right) \text { with }
   \Ar = 2 A, \Br = A^2 - 4 B; \\
\Kr   = \Q (\yr) & =  \Q [\Yr] / \left( (\Yr)^4 + \Ar (\Yr)^2 + \Br
\right).
\end {align*}
The minimal polynomials of $\yr$ over $K$ and $y$ over
$\Kr$ follow:
\begin{gather*}
    (\yr)^2 - 2 y \yr + (2 y^2 + A), \qquad
(y)^2 - \yr y + ((\yr)^2 + A)/2.
\end{gather*}
We write the Galois closure $L = K \Kr$ as the compositum generated
over~$K$ or~$\Kr$
by $t=y+\yr$. The minimal polynomial of $t$ is the resultant
\begin{align*}
    h (T) &= \Res_Y \left( Y^4+AY^2+B, (T-Y)^2 - 2 Y (T-Y) + (2 Y^2 + A)
\right)\\
    &= \Res_{\Yr} \left( (\Yr)^4 + \Ar (\Yr)^2 + \Br, (T-\Yr)^2 - \Yr
        (T-\Yr) +
    ((\yr)^2 + A)/2 \right)\\
    &= T^8 + 10AT^6 + (33A^2 - 14B)T^4 + (40A^3 - 70AB)T^2 +
    16A^4 - 200A^2B + 625B^2.
\end{align*}

\paragraph {Conversions and Galois actions.}

We are interested in the
action of~$\rho$, the generator of $\Gal (L / K)$, on~$\Kr$, and in the
action of~$\sigma$, the generator of $\Gal (L / \Kr)$, on~$K$. The
defining equations give:
\begin {eqnarray*}
&& \yr + (\yr)^\rho = 2 y, \quad
\yr (\yr)^\rho = \yr (\yr)^\rho = 2 y^2 + A, \quad
y^\rho = y,
\\
&& y +y^\sigma = \yr, \quad
y y^\sigma = \left( (\yr)^2 + A \right) / 2, \quad
(\yr)^\sigma = \yr.
\end {eqnarray*}

An element of $K$ is converted to an element of
$L$, as a relative extension of $\Kr$, using the identity $y=t-\yr$; in
the opposite direction we use $\yr=t-y$.
The entries of the matrix~$M$ of~\eqref {eq:period} are obtained from elements
of~$K$ and their images under~$\sigma$, and need to be expressed as elements
of~$\Kr$. For this we use the identity $y^\sigma = \yr-y$. 
This allows us to work in the relative extension $L / \Kr$ and to easily
identify elements of~$\Kr$.

\paragraph {Dual type norms.}
For an ideal $\bg$ of $\Kr$, we have
\[
\Norm_{\Phir} (\bg) = \Norm_{L / K} (\bg \Oc_L),
\]
see \cite[\S3.1]{BrGrLa11}. Computing dual type norms thus reduces to
conversions in relative extensions as described above.

\subsubsection {The cyclic case}

We may use the same type of equations for~$K$ and $K_0$ as in the dihedral
case, and may fix $\psi = \phi_1$ as in \eqref {eq:phiy}.
Fixing an arbitrary element $\sigma \in \Gal (K / \Q)$ of order~$4$,
we obtain $\phi_2 = \phi_1 \sigma$. Then the dual type norm for an ideal~$\bg$
of~$K$ is computed as
\[
\Norm_{\Phir} (\bg) = \bg \overline \bg^\sigma,
\]
see \cite[\S3.1]{BrGrLa11}.

\subsection {Symbolic reduction of period matrices}
\label{ssec:symbolic-reduction}

Gottschling in \cite{Gottschling59} has determined a finite set of
inequalities describing a fundamental domain $\Fc_2$ for
$\lquo {\Hc_2}{\Sp_4 (\Z)}$, which directly translate into an algorithm for
\textit {reducing} an element of $\Hc_2$ into $\Fc_2$. As the Igusa functions
introduced in \S\ref {ssec:igusa} are modular for $\Sp_4 (\Z)$, we may
transform all period matrices occurring in
Algorithm~\ref {algo:igusa-classpol} into~$\Fc_2$.
A period matrix~$\Omega$ is \textit {reduced} if $\Re (\Omega)$ has
coefficients between $-\frac {1}{2}$ and $\frac {1}{2}$ (which may be
obtained by reducing modulo~$\Z$), if the binary quadratic form defined
by $\Im (\Omega)$ is reduced (which
may be obtained using Gau{\ss}'s algorithm) and if
$|\det (C \Omega + D)| \geq 1$ for each of $19$ matrices
$\begin {pmatrix} A & B \\ C & D \end {pmatrix} \in \Sp_4 (\Z)$
(which may be obtained by applying to $\Omega$ a matrix for which the
condition is violated).
The process needs to be iterated and terminates eventually.

In the light of~\eqref {eq:period}, $\Omega = \psi (M)$ with
$M \in (\Kr)^{2 \times 2}$ and an explicitly given
$\psi : \Kr \to \C$, see~\eqref {eq:psiyp} and~\eqref {eq:phiy}.
Letting $\Kr = \Kr_0 + \yr \Kr_0$ as before, we have
$\psi|_{\Kr_0} : \Kr_0 = \Q (\sqrt {\Dr}) \to \R$ and
$\psi|_{\yr \Kr_0} : \yr \Kr_0 \to i \R$.
So $\Re (M)$ and $\Im (M)$ are the images under $\psi$ of matrices with
entries in~$\Kr_0$. The condition
$|\det (C \Omega + D)| \geq 1$ can be rewritten as
$\sqrt {\det (C \psi (M) + D)\det (C \psi (\overline M) + D)} \geq 1$
and thus also depends only on the images under $\psi$ of elements of~$\Kr_0$.

Hence the period matrices may be transformed symbolically into the fundamental
domain~$\Fc_2$ without computing complex approximations of their entries,
which precludes rounding errors:
The test whether the matrix is reduced and, if not,
the decision which transformation to apply depend on the sign of
$\psi (\alpha)$ for some $\alpha \in \Kr_0$, that is, on the sign
of some explicitly known $a + b \sqrt {\Dr} \in \R$, where $\sqrt {\Dr}$ is
the positive root of $\Dr$ and $a$, $b \in \Q$. This sign can be determined
from the signs of~$a$ and~$b$ and the relative magnitudes of~$a^2$
and~$b^2 \Dr$.

\section {Computing the Shimura group and its type norm subgroup}
\label{sec:shimuragroup}
\subsection{Structure of the Shimura group \texorpdfstring{$\Cf$}{C}}

The first step of Algorithm~\ref {algo:igusa-classpol} requires to enumerate
the Shimura group~$\Cf$ of~\eqref {eq:shimuragroup}, or more precisely, its
type norm subgroup $\Norm_{\Phir} (\Cl_{\Kr})$. We need the following exact
sequence, a proof of which can be found in \cite {BrGrLa11}:
\begin {equation}
\label {eq:exactshimura}
1\longrightarrow
\Oc_{K_0}^+/N_{K/K_0}(\Oc_K^*)
\xrightarrow{u\mapsto(\Oc_K,u)}
\Cf
\xrightarrow{(\af,\alpha)\mapsto\af}
\Cl_K
\xrightarrow{N_{K/K_0}}
\Cl^+_{K_0}
\longrightarrow
1,
\end {equation}
where $\Oc_{K_0}^+$ is the subgroup of totally positive units in $\Oc_{K_0}$
and $\Cl^+_{K_0}$ is the narrow class group of $K_0$.

We have algorithms at hand for the basic arithmetic of~$\Cf$.
For a finite abelian group, decomposed as a direct product of cyclic
groups~$G_i$ of order $d_i$ with $d_i \mid d_{i+1}$, we call the $d_i$
the \textit {elementary divisors} and a system of generators of the $G_i$
a \textit {(cyclic) basis} of the group.
Such a basis can be computed for the class group $\Cl_K$ (quickly under GRH)
using the function \code {bnfinit} in \parigp.
Equality testing of $(\af, \alpha)$ and $(\bg, \beta)$ amounts to testing
whether $\af \bg^{-1}$ is principal (either using \code {bnfisprincipal} in
\parigp, or by a direct comparison if each ideal is stored together with its
\textit {generalised discrete logarithm}, its coefficient vector with
respect to the basis of the class group), and whether
$\alpha / \beta = 1$ in $\Oc_{K_0}^+ / \Norm_{K / K_0} (\Oc_K^*)$.
Let $\epsilon_0$ and $\epsilon$ be the fundamental units of $K_0$ and $K$,
respectively.
If $\Norm (\epsilon_0) = -1$, then
$\Oc_{K_0}^+ = \langle \epsilon_0^2 \rangle
= \Norm_{K / K_0} (\langle \epsilon_0 \rangle)
\subseteq \Norm_{K / K_0} (\Oc_K^*)$,
and the quotient group is trivial.
If $\Norm (\epsilon_0) = +1$, then
$\Oc_{K_0}^+ = \langle \epsilon_0 \rangle$, and since
$\epsilon_0^2 = \Norm_{K / K_0} (\epsilon_0) \in \Norm_{K / K_0} (\Oc_K^*)$,
the quotient group is either trivial or
$\langle \epsilon_0 \rangle / \langle \epsilon_0^2 \rangle$,
in which case \code {bnfisunit} of \parigp\ can be used to compute the
exponent of the unit.

Multiplication is straightforward and can be made more efficient by a
\textit {reduction} step that outputs a smaller (not necessarily unique)
representative. To reduce $(\af, \alpha)$, one computes an LLL-reduced
ideal $\af' = \mu \af$ (using \code {idealred} in \parigp) and
lets $\alpha' = \mu \overline \mu \alpha$
One then tries to reduce the unit contribution in the size of the
algebraic number $\alpha'$ by multiplying it with an appropriate power
of $\Norm_{K / K_0} (\epsilon)$.

The Shimura group $\Cf$ and its subgroup $\Norm_{\Phir} (\Cl_{K_r})$ can be
enumerated directly; but the map $\Norm_{\Phir} : \Cl_{K_r} \to \Cf$ being
in general non-injective, this can require a large number of expensive
principality tests in $\Cf$ to avoid duplicates. More elegantly, we may
consider the groups in~\eqref {eq:exactshimura} as given by cyclic bases or,
more generally, generators and relations, and complete the sequence from
known data using tools of linear algebra for $\Z$-modules, in particular the
Hermite (HNF) and Smith normal forms (SNF), see \cite[\S2.4]{Cohen93}.

\begin {algorithm}
\label{algo:ck-structure}
\inoutput {Cyclic bases for $\Cl_K$ and $\Cl^+_{K_0}$}
{Cyclic basis for $\Cf$}
\begin {enumerate}
\item Compute a matrix $M$ for $\Norm_{K / K_0} : \Cl_K \to \Cl^+_{K_0}$.
\item Compute generators $\af_1, \ldots, \af_r$ of the kernel of $M$.
\item Lift $\af_1, \ldots, \af_r$ to $\Cf$: Pick arbitrary totally positive
   $\alpha_i \in K_0$ such that $\af_i \overline \af_i = \alpha_i \Oc_{K_0}$.
\item Compute a basis for the lattice $L_0$ of relations such that
   the subgroup of $\Cl_K$ generated by $\af_1, \ldots, \af_r$ is
   isomorphic to $\Z^r / L_0$.
\item If $\Oc^+_{K_0} / \Norm_{K / K_0} (\Oc^*) = 1$, let $r' = r$;
   otherwise let $r' = r+1$ and
   $(\af_{r'}, \alpha_{r'}) = (\Oc_K, \epsilon_0)$.
\item Expand the basis of \step {4} into a basis for the lattice~$L$ of
   relations between the generators
   $(\af_1, \alpha_1), \ldots, (\af_{r'}, \alpha_{r'})$
   such that $\Cf \simeq \Z^{r'} / L$.
\item Determine a cyclic basis of $\Cf$.
\end {enumerate}
\end{algorithm}

Step~\step {1} requires to apply the generalised discrete logarithm map in
$\Cl^+_{K_0}$ to the small number of relative norms of the basis elements
of~$\Cl_K$.
Step~\step {3} is possible since the
$\af_i \overline \af_i = \Norm_{K / K_0} (\af_i)$
are trivial in~$\Cl^+_{K_0}$.
Steps~\step {5} and~\step {6} rely on the exactness of the
sequence~\eqref {eq:exactshimura}. If $r' = r$, there is nothing to do.
Otherwise, we first add the relation $(\Oc_K, \epsilon_0)^2 = 1$.
Lifts of relations from~$L_0$ are then in the image of
$\langle \epsilon_0 \rangle / \langle \epsilon_0^2 \rangle$,
and if the unit exponent is odd in the lift, we need to add
$(\Oc_K, \epsilon_0)$ into the relation.
Steps~\step {2} and~\step {4} require an~HNF, Step~\step {7} an~SNF.

\subsection{The type norm subgroup}
\label {sec:typenorm}

Algorithm~\ref {algo:ck-structure} also provides an algorithm for generalised
discrete logarithms in~$\Cf$, which can be used to determine the subgroup
$N_{\Phir} (\Cl_{\Kr})$ in a similar way:
For each generator of~$\Cl_{K^r}$, we compute the generalised discrete
logarithm of its image in $\Cf$, then the relations between the images
using an HNF and a cyclic basis using an SNF.
The enumeration of the subgroup is then trivial.
In the same vein, it is possible to compute all the cosets
$\Cf / N_{\Phir} (\Cl_{\Kr})$ if the complete Igusa class polynomial
is desired and not only its irreducible factor $H_1$,
see \S\ref {ssec:igusa}.

\section {Computing \texorpdfstring{$\theta$}{Θ}-constants}
\label {sec:computing-thetas}

As explained in Step~\step {4} of Algorithm~\ref {algo:igusa-classpol},
it suffices to compute the fundamental $\theta$-constants
$\theta_0, \ldots, \theta_3$ in the argument $\Omega / 2$
to obtain the class invariants for the period
matrix~$\Omega =
\begin {pmatrix} \omega_0 & \omega_1 \\ \omega_2 & \omega_0 \end {pmatrix}
\in \Fc_2$.

In \S\ref {ssec:theta-naive} we describe an algorithm to compute the
$\theta$-constants directly from their $q$-expansions, using a lower
number of multiplications than approaches described previously
in the literature.

As the coefficients of the Igusa class polynomials grow rather quickly,
a high floating point precision is needed for evaluating the
$\theta$-constants. In \S\S\ref {ssec:borchardt}--\ref {ssec:theta-newton}
we describe an algorithm with a quasi-linear (up to logarithmic factors)
complexity in the desired precision, using Newton iterations on a function
involving the Borchardt mean. The algorithm is described essentially in
Dupont's PhD thesis~\cite{Dupont06}; for the corresponding algorithm
in dimension~$1$, using the arithmetic-geometric mean instead of the
Borchardt mean, see \cite{Dupont11}. We provide a streamlined presentation
in dimension~$2$, together with improved algorithms and justifications.

\subsection{Naive approach}
\label{ssec:theta-naive}

For the fundamental $\theta$-constants, \eqref {eq:theta} specialises as
\begin {equation}
\label {eq:theta-fundamental}
\theta_{4 b_1 + 2 b_2} (\Omega / 2)
= \sum_{m, n \in \Z} (-1)^{2 (m b_1 + n b_2)} q_0^{m^2}q_1^{2mn}q_2^{n^2}
\end {equation}
with $q_k = \exp(i\pi\omega_k/2)$.

Positive definiteness and reducedness of the binary quadratic form
attached to $\Im (\Omega)$ show that the sum converges when taken
over, for instance, a square $[-R, R]^2$ with $R \to \infty$;
\cite[p. 210 following the proof of Lemma~10.1, with typos]{Dupont06}
establishes that for $R \geq \sqrt {1.02 N + 5.43}$, the truncated sum is
accurate to $N$ bits.
Better bounds may be reached using summation areas related to the
eigenvalues of $\Im (\Omega)$, but using a square allows to organise
and reuse computations so as to reduce the number of multiplications
of complex numbers.

\begin {proposition}
\label {prop:naive}
The truncated sum over $(m, n) \in [-R, R]^2$ for the fundamental
$\theta$-constants \eqref {eq:theta-fundamental} may be computed with
$2 R^2 + O (R)$ multiplications and one inversion
using storage for $R + O (1)$ elements.

Letting $R = \lceil \sqrt {1.02 N + 5.43} \rceil$ and using complex numbers
of precision ${O}(N)$, we obtain a time complexity of
\[
O(N \mult (N))
\text { or }
\Ot(N^2),
\]
where
$\Ot(N) = O\left(N(\log N)^{O(1)}\right)$,
and $\mult (N) \in \Ot(N)$ is the time complexity of multiplying two
numbers of $N$ bits.
\end {proposition}

\begin {proof}
Using symmetries with respect to the signs of $m$ and $n$, we may write
\begin {multline*}
\sum_{-R \leq m, n \leq R} (-1)^{2 (m b_1 + n b_2)} q_0^{m^2}q_1^{2mn}q_2^{n^2}
=
1
+ 2 \sum_{m = 1}^R (-1)^{2 m b_1} q_0^{m^2}
+ 2 \sum_{n = 1}^R (-1)^{2 n b_2} q_2^{n^2} \\
+ 2 \sum_{m = 1}^R (-1)^{2 m b_1} q_0^{m_2}
    \sum_{n = 1}^R (-1)^{2 n b_2} q_2^{n^2}
                                  \left( q_1^{2 m n} + q_1^{- 2 m n} \right).
\end {multline*}

We first compute and store the $q_2^{n^2}$ with $2 R + O (1)$ multiplications
via $q_2^{2 n - 1} = q_2^{2 (n-1) - 1} \cdot q_2^2$ and
$q_2^{n^2} = q_2^{(n-1)^2} \cdot q_2^{2 n - 1}$.
After computing the inverse $q_1^{-1}$, a similar scheme yields the
$q_0^{m^2}$ and $q_1^{2 m} + q_1^{-2 m}$ without storing them.
At the same time, we may compute for any given~$m$ the sum over~$n$
inside the double sum:
The term $q_1^{2 m n} + q_1^{- 2 m n}$ is the $n$-th element $v_n$ of the
Lucas sequence $v_0 = 2$, $v_1 = q_1^{2 m} + q_1^{-2 m}$,
$v_n = v_1 \cdot v_{n-1} - v_{n-2}$, each element of which is computed with
one multiplication. Together with the multiplication by $q_2^{n^2}$, each
term of the innermost sum is thus obtained with two multiplications.

For the time complexity, recall that complex inversions can be computed
in time $O (\mult (N))$, and exponentials in time $O (\mult (N) \log N)$,
see \cite {Brent76}.
\end {proof}

This algorithm gains an asymptotic factor of $2/3$ over
\cite [Algorithme~15]{Dupont06}.

\subsection{Borchardt mean of complex numbers}
\label {ssec:borchardt}

The key tool in the asymptotically fast evaluation of $\theta$-constants
is the Borchardt mean, which generalises Lagrange's and Gauß's
arithmetic-geometric mean of two numbers to four.
The Borchardt mean of four positive real numbers has been introduced
in~\cite{Borchardt76,Borchardt78}. The complex case is treated in
\cite{Dupont06}, where proofs of most (but not all) propositions below
may be found.
It is made complicated by the presence of several square roots in the
formulæ, each of which is defined only up to sign.

\begin {definition}
Let
\begin {eqnarray*}
\Hc
& = & \left\{ z \in \C :
   \arg (z) \in \left] - \frac {\pi}{2}, \frac {\pi}{2} \right] \right\}
   \cup \{ 0 \} \\
& = & \{ z \in \C : \Re (z) > 0, \text { or } \Re (z) = 0 \text { and }
\Im (z) \geq 0 \}
\end {eqnarray*}
be the \emph {complex half-plane} defining the standard branch of the
complex square root function.
For a number in $\Hc$, its square root in $\Hc$ lies in fact in the
\emph {complex quarter-plane}
\[
\Qc = \left\{ z \in \C :
   \arg (z) \in \left] - \frac {\pi}{4}, \frac {\pi}{4} \right] \right\}
   \cup \{ 0 \}.
\]
\end {definition}

\begin {defprop}
\label {def:borchardt}
Given a complex quadruple
$b = (b_0, \ldots, b_3) \in \C^4$,
a \emph {Borchardt iterate}
is a quadruple
$b' = (b'_0, \ldots, b'_4)$ such that there are four choices of square roots
$(\sqrt{b_j})_{j=0,\ldots, 3}$ yielding
\begin{align*}
    b'_0 &= \tfrac14(b_0+b_1+b_2+b_3)&
    b'_1 &= \tfrac12(\sqrt{b_0}\sqrt{b_1}+\sqrt{b_2}\sqrt{b_3})\\
    b'_2 &= \tfrac12(\sqrt{b_0}\sqrt{b_2}+\sqrt{b_1}\sqrt{b_3})&
    b'_3 &= \tfrac12(\sqrt{b_0}\sqrt{b_3}+\sqrt{b_1}\sqrt{b_2})
\end{align*}
There are up to eight different Borchardt iterates of a given quadruple.
If $b \in \Hc^4$, the \emph {standard Borchardt iterate} is obtained by
choosing square roots in $\Qc$, so that $b' \in \Hc^4$ again.
More generally, if all entries of $b$ lie in the same half-plane, that is,
$b \in (z \Hc)^4$ for some $z \in \C$, choosing all square roots in the same
quarter-plane $\sqrt z \Qc$ (with either choice of sign for $\sqrt z$)
yields the standard Borchardt iterate in the same half-plane.

A \emph {Borchardt sequence} is a sequence $\left( b^{(n)} \right)_{n\geq0}$
such that $b^{(n+1)}$ is a Borchardt iterate of $b^{(n)}$ for all $n\geq0$.
If all entries of $b^{(0)}$ lie in the same half-plane, its
\emph {standard Borchardt sequence} is defined by taking only standard
Borchardt iterates.
\end{defprop}

The following result is proved in~\cite[Chapter 7]{Dupont06}.
\begin{proposition}
\label{prop:borchardt}
Any Borchardt sequence converges to a limit $(z, z, z, z)$.

When the elements of $b$ are contained in the same half-plane,
the \emph {Borchardt mean} $B_2 (b)$ of~$b$ is the limit of the standard
Borchardt sequence starting with $b^{(0)} = b$. The function $B_2$ is
obviously homogeneous.

A standard Borchardt sequence converges quadratically:
\[
\left\lVert b^{(n)} - B_2 (b) = 2^{-O(2^n)} \right\rVert.
\]
This implies that the Borchardt mean is computed to a precision of $N$ bits
with $O (\log N)$ multiplications in time
\[
O (\mult (N)\log N).
\]
\end{proposition}

Comparison of the formulæ in Definition~\ref {def:borchardt}
and~\eqref {eq:duplication} shows that for any period matrix
$\Omega\in\Hc_2$, the sequence
$\left( (\theta^2_j(2^n\Omega))_{j=0,\ldots, 3} \right)_{n \geq 0}$
is a Borchardt sequence. This fact alone does not solve the sign issue,
however. One would hope for the $\theta$-sequence to be the standard
Borchardt sequence, which would allow it to be computed with the standard
choice of complex square roots. This assumption does not hold in general;
however, it is true for the fundamental $\theta$-constants and
$\Omega \in \Fc_2$.

\begin{proposition}
\label{prop:b2-theta-convergence-id}
For $\Omega\in\Fc_2$, $n \geq 0$ and $j = 0, \ldots, 3$
we have
$\theta_j(2^n\Omega) \in \Qc$.
Hence $\left( (\theta^2_j(2^n\Omega))_{j=0,\ldots, 3} \right)_{n\geq0}$ is
the standard Borchardt sequence associated to
$(\theta^2_j(\Omega))_{j=0,\ldots, 3}$.
It converges to $1$.
\end{proposition}
The result follows from \cite[Propositions 6.1 and 9.1]{Dupont06}.

\subsection {Period matrix coefficients from \texorpdfstring{$\theta$}{Θ}-constants}

For the time being, we consider the inverse of the function we are interested
in and describe an algorithm that upon input of the values of the four
fundamental $\theta$-quotients in a period matrix returns the coefficients of the
period matrix. Newton iterations can then be used to invert this function.

By the modularity of the squares of the $\theta$-constants, applying a
matrix $\gamma \in \Sp_4 (\Z)$ to their argument $\Omega$ permutes the
functions and multiplies them by a common projective factor, which depends
on $\gamma$ and $\Omega$. In this way, information on $\Omega$ can be
gathered; informally, three matrices suffice to obtain the three different
coefficients of $\Omega$. We consider three particular matrices, as
suggested in \cite [\S9.2.3]{Dupont06}, which lead to well-behaved Borchardt means,
see Conjecture~\ref {conj91}.

\begin{proposition}
\label {prop:periodfromborchardt}
    Let $\mathfrak{J}=
    \begin{pmatrix}0&-\id_2\\\id_2&0\end{pmatrix}$ and $\mathfrak{M}_j=
\begin{pmatrix}\id_2&m_j\\0&\id_2\end{pmatrix}$ with
    $m_0=\begin{pmatrix}1&0\\0&0\end{pmatrix}$,
    $m_1=\begin{pmatrix}0&1\\1&0\end{pmatrix}$,
    $m_2=\begin{pmatrix}0&0\\0&1\end{pmatrix}$.
    Let $\Omega\in\Hc_2$.
    Then
    \begin{align*}
\left( \theta^2_j((\mathfrak{J}\mathfrak{M}_0)^2 \Omega) \right)_{j=0,1,2,3}
&= -i\omega_0 \left( \theta^2_j(\Omega) \right)_{j=4,0,6,2},\\
\left( \theta^2_j((\mathfrak{J}\mathfrak{M}_1)^2 \Omega) \right)_{j=0,1,2,3}
&= (\omega_1^2-\omega_0\omega_2) \left( \theta^2_j(\Omega) \right)_{j=0,8,4,12},\\
\left( \theta^2_j((\mathfrak{J}\mathfrak{M}_2)^2 \Omega) \right)_{j=0,1,2,3}
&= -i\omega_2 \left( \theta^2_j(\Omega) \right)_{j=8,9,0,1}.
    \end{align*}
\end{proposition}
A more general statement with the action on the $\theta$-constants (not
squared) is given in~\cite[Propriété 3.1.24]{Cosset11},
following~\cite[Chapter 5, Theorem 2]{Igusa72}.  The explicit form
restricted to squares of $\theta$-constants, as given here, is found
in~\cite[\S6.3.1]{Dupont06}.

The idea of the algorithm is now to apply the Borchardt mean function
$B_2$ to both sides of the above equations. Conjecturally, the left
hand side becomes~$1$, so that each Borchardt mean of a right hand side
yields a coefficient of $\Omega$. So we rely on the following conjecture,
for which we have overwhelming numerical evidence, but no complete proof.
Notice that it is \textit {a priori} not even clear if the Borchardt
means are well-defined, that is, if the squares of the various four
$\theta$-values always lie in the same half-plane.

\begin{conjecture}
\label{conj91}
Let
\[
\Uc = \left\{ \Omega\in\Hc_2 :
    B_2 \left( (\theta^2_j(\Omega))_{j=0,\ldots, 3} \right)
    \text{ is defined and equal to $1$} \right\}.
\]
For $k\in\{0,1,2\}$ we have
$\displaystyle
(\mathfrak{J}\mathfrak{M}_k)^2 \Fc_2\subseteq\mathcal{U}$.
\end{conjecture}

Under Conjecture~\ref {conj91}, we can now formulate an algorithm to obtain
$\Omega$ from four values of $\theta$-constants. To make the following
Newton iterations more efficient, we dehomogenise all modular functions
by dividing by appropriate powers of $\theta_0$, which allows to work
with only three inputs.

\begin {algorithm}
\label{algo:tau_from_bj}
\inoutput {Floating point approximations of
    $\left( \theta_j(\Omega/2)/\theta_0(\Omega/2) \right)_{j=1,2,3}$ for some
$\Omega\in\Fc_2$, and as auxiliary data the sign of $\omega_1$.}
{Floating point approximations of the coefficients $\omega_0$, $\omega_1$,
$\omega_2$ of $\Omega \in \Fc_2$}
\begin {enumerate}
\item Use the duplication formulæ \eqref {eq:duplication} to
compute $(\theta^2_j(\Omega)/\theta^2_0(\Omega/2))_{j=0,1,2,3,4,6,8,9,12,15}$.
\item Compute
        $B_2((\theta^2_j(\Omega)/\theta^2_0(\Omega/2))_{j=0,1,2,3})
        =\frac1{\theta^2_0(\Omega/2)}$.
\item Deduce $(\theta^2_j(\Omega))_{j=0,1,2,3,4,6,8,9,12,15}$.
\item Compute
    \begin{gather*}
        u_0=B_2((\theta^2_j(\Omega))_{j=4,0,6,2}), \quad
        u_2=B_2((\theta^2_j(\Omega))_{j=8,9,0,1}), \quad
        u_1=B_2((\theta^2_j(\Omega))_{j=0,8,4,12}).
    \end{gather*}
\item Return $\omega'_0=\frac{i}{u_0}$, $\omega'_2=\frac{i}{u_2}$ and
    $\omega'_1=\pm \sqrt{\frac1{u_1}+\omega'_0\omega'_2}$ with the
    appropriate sign.
\end{enumerate}
\end{algorithm}

The correctness of Algorithm~\ref{algo:tau_from_bj} under
Conjecture~\ref {conj91} follows from the discussions above.
Step~\step {1} uses the homogeneity of \eqref {eq:duplication},
Step~\step {2} the homogeneity of the Borchardt mean and
Proposition~\ref {prop:b2-theta-convergence-id}.
The validity of Step~\step {5} follows from
Proposition~\ref {prop:periodfromborchardt} under
Conjecture~\ref {conj91}, using again the homogeneity of the
Borchardt mean.

Notice that $\Omega$ is only well-defined up to the subgroup of
$\Sp_4 (\Z)$ for which the $\theta$-constants are modular.
Assuming $\Omega \in \Fc_2$, the fundamental domain for all of
$\Sp_4 (\Z)$, it is necessarily unique; Conjecture~\ref {conj91}
implies that this particular representative for $\Omega$ is indeed
returned by the algorithm.

\subsection{Newton lift for fundamental
\texorpdfstring {$\theta$}{theta}-constants}
\label {ssec:theta-newton}

Denote by
\[
F : \C^3 \to \C^3, \quad
\left( \theta_j(\Omega/2)/\theta_0(\Omega/2) \right)_{j=1,2,3} \mapsto \Omega,
\]
the function computed by Algorithm~\ref {algo:tau_from_bj},
and by
\[
f : \Fc_2 \to \C^3, \quad
\Omega \mapsto \left( \theta_j(\Omega/2)/\theta_0(\Omega/2) \right)_{j=1,2,3},
\]
its inverse on $\Fc_2$
(where $\Omega$ is interpreted as the three-element vector
$(\omega_0, \omega_1, \omega_2)$ and not as a four-element matrix).

We use Newton iterations on~$F$ to compute~$f$.
The standard Newton approach requires to compute the Jacobian matrix~$J_F$
of~$F$, that is, its partial derivatives with respect to its different
coordinates. Heuristically, Algorithm~\ref {algo:tau_from_bj} may be modified
accordingly to also output~$J_F$, see \cite[Algorithme~16]{Dupont06},
generalising the dimension~$1$ approach of~\cite[\S2.4]{BoBo87} and
\cite {Dupont11}. The description and justification of this algorithm
are rather technical.
Instead, we opt for using finite differences, which moreover turn out to
yield a more efficient algorithm (see~\S\ref{ssec:impl-theta}).

\begin {algorithm}
\label {algo:newton}
\inoutput {Floating point approximations $y^{(n)}$ of $\Omega \in \Fc_2$,
    to precision $2N$ bits,
and $x^{(n)}$ of $f (\Omega)$, 
to precision $N$ bits.}
{Floating point approximation $x^{(n+1)}$ of $f (\Omega)$,
to precision $2N$ bits (see Theorem~\ref {th:newton}).}
\begin {enumerate}
\item
    Let $\epsilon = 2^{-N}\max_j\left\{ \left| x_j^{(n)} \right| \right\}$.
\item
Let $(e_j)_{j=1, 2, 3}$ be the standard basis of $\C^3$.
By Algorithm~\ref {algo:tau_from_bj}, compute
$F (x^{(n)})$ and $\frac{\Delta F}{\Delta x_j} =
\frac{1}{\epsilon}\left(F (x^{(n)} + \epsilon e_j) -
F(x^{(n)})\right)$.
\item
Let $J = (J_{ij})_{i, j = 1, 2, 3}$ with
$J_{ij} = \frac{\Delta F_i}{\Delta x_j}$.
\item
Let
\[
x^{(n+1)} = x^{(n)} - \left( F (x^{(n)}) - y^{(n)} \right) J^{-1}
\]
(where all vectors are seen as row vectors).
\end {enumerate}
\end {algorithm}

All computations in the algorithm are carried out at a precision of
$2 N$ bits. But even without taking rounding errors into account,
the approximation of the Jacobian matrix by finite differences as well
as the Newton method itself introduce some inaccuracy in the result,
so that the accuracy improves to less than $2 N$ bits. The
following proposition addresses this issue.

\begin{theorem}
\label{th:newton}
Assume the validity of Conjecture~\ref {conj91}.
Let $\Omega \in \Fc_2$ be such that $\theta_0 (\Omega / 2) \neq 0$,
$x = f (\Omega)$, and let $x^{(0)}$ be an initial floating point
approximation to~$x$.
Not taking rounding errors into account,
there exist two real numbers
$\epsilon_0>0$ and $\delta>0$, depending on $x$, such that
for $\lVert x^{(0)} - x \rVert < \epsilon_0$, the sequence $x^{(n)}$
defined by successive applications of Algorithm~\ref {algo:newton}
converges to~$x$,
with accuracy increasing in each step from $N$ to $2 N - \delta$.

To reach a given accuracy~$N$, the total complexity is dominated by the
complexity of the last lifting step, that is:
\[
O (\mult (N) \log N).
\]
\end {theorem}

\begin {proof}
By assumption, $F$ is defined and analytic in a neighbourhood of~$x$.
In particular, its second partial derivatives are bounded
in a close enough neighbourhood of~$x$, so that
the Jacobian matrix of $F$ in $x^{(n)}$ is approximated
to accuracy $2 N-\delta_0$ bits by the matrix~$J$ computed in Steps~\step {2}
and~\step {3}, where $\delta_0$ depends on $x$ and on the bound on the
second partial derivatives.
The remaining assertion, with some $\delta \geq \delta_0$,
is the standard result for Newton's
method (see~\cite[Chapter 9 and §15.4]{GaGe99} and~\cite[§4.2]{BrZi10}).

The complexity is derived from the superlinearity of multiplication,
which makes the last of the $O (\log N)$ Newton steps dominate the whole
computation; the logarithmic factor stems from the complexity of computing
the Borchardt mean given in Proposition~\ref {prop:borchardt}.
\end {proof}

Notice that for our application of computing class polynomials for primitive
quartic CM fields, the assumption of Theorem~\ref {th:newton} is satisfied:
As $\begin {pmatrix} \Omega & \id_2 \end {pmatrix} \Z^4$ is of rank~$4$,
we have $\omega_1 \neq 0$,
and therefore none of the $\theta_j(\Omega/2)$ vanish
(see~\cite[Chapter~9, Proposition~2]{Klingen90}).

In practice, we use a fixed initial precision for $x^{(0)}$, computed
according to Proposition~\ref {prop:naive}, which determines $\epsilon$
and implicitly $\delta$. The lack of information about $\delta$
can be worked around as follows:
If $x^{(n-2)}$ and $x^{(n-1)}$ agree to~$k$ bits, and
$x^{(n-1)}$ and~$x^{(n)}$ agree to~$k'$ bits, we set $\delta=2k-k'$.
This value of $\delta$ accounts at the same time for bits lost due to
rounding errors induced by the floating point computations.

\begin {remark}
It is possible to modify Algorithm~\ref {algo:tau_from_bj} and consequently
Algorithm~\ref {algo:newton} so as not to rely on Conjecture~\ref {conj91}.
The conjecture states that the choices of square roots inside the Borchardt
mean computations correspond to doubling the argument of the
$\theta$-constants.
So by computing very low precision approximations of the $\theta$-constants
in $2^n \Omega$ as described in \S\ref {ssec:theta-naive}, one can make
sure to choose the correct sign. These computations do not deteriorate the
asymptotic complexity; moreover, as Algorithm~\ref {algo:newton} requires
the Borchardt means of the same arguments over and over again (albeit with
increasing precision), the sign choices may be fixed once and for all in a
precomputation step.

In practice, however, we did not come upon any counterexample to
Conjecture~\ref {conj91} with tens of thousands of arguments.
\end {remark}

\section {Reconstruction of class polynomial coefficients and
reduction modulo prime ideals}

\label {sec:lll}

\subsection {The dihedral case}

The class polynomials $H_1$, $\Hhat_2$, $\Hhat_3$ of~\eqref {eq:H}
and~\eqref {eq:Hhat} for a fixed CM type~$\Phi$ are defined over~$\Kr_0$,
but Steps~\step {1} to~\step {5} of Algorithm~\ref {algo:igusa-classpol}
compute floating point approximations, precisely of the images of the
polynomials under an embedding $\psi : \Kr_0 \to \C$.
To realise Step~\step {6} of Algorithm~\ref {algo:igusa-classpol}, we
need to invert $\psi$: Given $\psi (\alpha)$ to sufficient precision, we
wish to reconstruct $\alpha$ symbolically as an element of
$\Kr_0 = \Q (\zr) = \Q [\Zr] / \left( (\Zr)^2 + \Ar \Zr + \Br \right)$,
cf.~\S\ref {ssec:nf}.
We may limit the discussion to the CM type $\Phi$ and the embedding~$\psi$
given by~\eqref {eq:psiyp} and~\eqref {eq:phiy}; the second CM type $\Phi'$
leads to class polynomials that are conjugate under $\Gal (\Kr_0 / \Q)$.

Let $\Dr$ be the discriminant of $\Kr_0$; as the discriminant of
the minimal polynomial of $\zr$ is $16 B$, we have $\Kr_0 = \Q (\sqrt B)$,
and $\frac {\Dr}{B}$ is a rational square.
Let $w \in \Kr_0$ with $w^2 = \Dr$ satisfy $\psi (w) = \sqrt {\Dr} > 0$.
Write $\alpha = \frac {a + b w}{c}$ with coprime $a$, $b$, $c \in \Z$.
Knowing an approximation $\beta$ to $\psi (\alpha) \in \R$ at our working
precision of $n$~bits, we wish to recover $a$, $b$, $c$, for which there is
hope if $2^n > |a b c|$.

Let $e$ be the exponent of $\beta$ in the sense that
$2^{e-1} \leq |\beta| < 2^e$, and let
$e^+ = \max (e, 0)$ and $e^- = \max (-e, 0)$, so that $e = e^+ - e^-$,
and at most one of $e^+$, $e^-$ is non-zero.
We expect $|a| \approx \sqrt {\Dr}\, |b|$ (whereas $c$ is usually smaller),
so that $2^{e^-} |a| \approx 2^{e^-} \sqrt {\Dr} \, |b|
\approx 2^{e^-} |c \beta| \approx 2^{e^+} |c|$.
On the other hand, the floating point approximation $\beta$ satisfies
$\left| \beta - \frac {a + b \sqrt {\Dr}}{c} \right|
\approx 2^{e - n}$
(up to a small factor accounting for digits lost to rounding errors), whence
$2^{n + e^-} \left| c \beta - (a + b \sqrt {\Dr}) \right| \approx 2^{e^+} c$
is comparative in size to the previous quantities.
Consider the integral matrix
\[
\begin {pmatrix}
0 & 0 & 2^{n + e^+} \\
\left\lfloor 2^{e^-} \sqrt {\Dr} \right\rceil & 0
  & \left\lfloor 2^{n + e^+} \sqrt {\Dr} \right\rceil \\
0 & 2^{e^+} & \left\lfloor \beta 2^{n + e^+} \right\rceil
\end {pmatrix}.
\]
Using LLL, we find a short vector
$\left( -b \left\lfloor 2^{e^-} \sqrt {\Dr} \right\rceil,
c 2^{e^+}, r \right)$
in the lattice spanned by the rows of the matrix; the scaling of the last
column was chosen, following the arguments above, such that all entries in
the vector have comparable sizes.
This determines $b$ and~$c$, and we let
$a = \frac {c \left\lfloor \beta 2^{n + e^+} \right\rceil
- b \left\lfloor 2^{n + e^+} \sqrt {\Dr} \right\rceil}{2^{n + e^+}}
\in \Z$.

To get back to our standard representation of $\Kr_0$, we need to relate
$w$ and $\zr$.
By \eqref {eq:psiyp} and \eqref {eq:phiy},
\[
\psi (\zr) = \psi (\yr)^2 = - A - 2 \sqrt B < 0,
\]
so that
\begin {equation}
\label {eq:w}
w = \sqrt {\frac {\Dr}{B}} \cdot \frac {-\zr - A}{2}.
\end {equation}

To obtain abelian varieties over finite fields, we need to reduce the
class polynomials modulo certain prime ideals $\pf_1$ of $\Kr_0$.
Let $p$ be a rational prime that splits as $\pf_1 \pf_2$ in $\Kr_0$.
Assume that $\pf_1$ splits in $\Kr$, so that $\pf_1 = \qf_1 \overline \qf_1$,
and that the type norm of $\qf_1$ is a principal ideal of $K$.
Then the class polynomial splits totally modulo $\pf_1$, and its reduction
may be computed as follows: If
$\pf_1 = p \Oc_{\Kr_0} + (a + b w) \Oc_{\Kr_0}$ with $a$, $b \in \Z$,
replace each occurrence of $w$ by $- \frac {a}{b}$ and reduce modulo~$p$.

\subsection {The cyclic case}

Here the class polynomials are defined over~$\Q$, its coefficients may be
obtained by a $2$-dimensional lattice reduction, and reduction modulo
primes is trivial.

\section {Implementation and parallelisation}
\label {sec:implementation}

Our implementation of the algorithms defined here is available in the
software package \cmh\cite {cmh}, which can be downloaded from
\begin{center}
\url{http://cmh.gforge.inria.fr/}.
\end{center}
The current version of the \cmh\ software package is still in
development, and will be named \cmh-1.0 once some packaging improvements,
alongside with minor bug corrections, are checked in.

The software implements the different steps of
Algorithm~\ref{algo:igusa-classpol} as follows:
\begin{itemize}
    \item Steps~\step1 to \step3 of Algorithm~\ref{algo:igusa-classpol} are
        performed by a script in \parigp\cite {parigp}, which does all
        computations symbolically, and the running time of which is
        essentially negligible.
    \item The computation of $\theta$-constants in Step~4) of
        Algorithm~\ref{algo:igusa-classpol} is done by a C
        program, based on the library \mpc\cite {mpc}, itself using the
        \mpfr\cite {mpfr} and \gmp\cite {gmp} libraries. Newton lifting is
        used for this step from a base precision of $2\,000$~bits,
        and it is parallelised through MPI.
    \item Reconstruction of the class polynomials from the numerical values
        of the Igusa invariants is done inside the same C program, relying
        on the library \mpfrcx\cite {mpfrcx} for basic operations
        on polynomials using the FFT and asymptotically fast algorithms on
        trees of polynomials.
In a preparatory step, the leaves of the tree for $H_1$ are filled
with the linear factors of the class polynomial, those for $\Hhat_k$,
$k = 2, 3$,
are filled with the values of $j_k$. Let the subscripts
$\mathrm l$ and $\mathrm r$ denote the left and the right descendant,
respectively, of a given node. Then an inner node $n^{(1)}$ in the tree
for~$H_1$ is computed as
$n^{(1)} = n^{(1)}_{\mathrm l} \cdot n^{(1)}_{\mathrm r}$,
while an inner  node $n^{(k)}$ in the tree for $\Hhat_k$, $k = 2, 3$,
is obtained as
$  n^{(k)}_{\mathrm l} \cdot n^{(1)}_{\mathrm r}
 + n^{(1)}_{\mathrm l} \cdot n^{(k)}_{\mathrm r}$,
where $n^{(1)}$ denotes the node at the same position in the tree for~$H_1$;
for details, see \cite [Algorithms~10.3 and~10.9]{GaGe99}.
By first combining pairs of complex-conjugate leaves in a preprocessing
step, all computations are in fact carried out with real floating point
polynomials, see \cite {EnMo03}.
So if at a given level the tree for~$H_1$ contains $m$ nodes, all nodes at
this level of the three trees can be obtained with $5 m$ independent
multiplications, which are parallelised using MPI.
    \item Recognition of the polynomial
        coefficients as elements in $K_0^r$ is also embedded in the
        C program, using \fplll\cite {fplll} for the LLL step.
    \item Assuming that $K \neq \Q (\zeta_5)$, for which the result is known,
        validation of the obtained class polynomials is performed by
        computing a Weil number $\pi$ above a prime $p\approx 2^{128}$,
        constructing a curve over $\F_p$ having as endomorphism ring the
        ring of integers of $K$
        using Mestre's algorithm~\cite{Mestre91}, and verifying
        that the cardinality of the Jacobian matches
        $\Norm_{K/\Q}(1\pm\pi)$. This step is done in \parigp\ and also
        has a negligible cost.
\end{itemize}

In the following we report on the performance of these
different steps, illustrated by both small and large examples.

\subsection{Computation of \texorpdfstring{$\theta$}{Θ}-constants}
\label{ssec:impl-theta}

We report timing results for the computation of fundamental
$\theta$-constants for two
arbitrary period matrices. Table~\ref{tab:theta-vs-magma} shows that already
our implementation of the
relatively simple naive algorithm presented in~\S\ref{ssec:theta-naive} may be
several orders of magnitude faster\footnote{Such a quadratic, yet efficient
implementation was used by T. Houtmann to compute class
polynomials of degree up to~500 (personal communication, no reference
exists).} than \magma-2.19.4, the performance improvement ratio
depending on the period matrix.
Newton lifting is preferable above some cut-off
value for the precision, here $16\,000$ and $4\,000$ bits, respectively.
The naive algorithm is rather sensitive to the period matrix; generally
speaking, it converges the faster the larger the imaginary parts in
$\Omega$ are, which correspond to smaller $q_0$, $q_1$, $q_2$.
A noticeable difference
between our naive algorithm from~\S\ref{ssec:theta-naive} and the
implementation in \magma\ is that the favorable cases are not the same.
This is most likely due do different choices of summation
regions, as briefly discussed in~\S\ref{ssec:theta-naive}.
We note that the timings of Newton lifting depend much less on the 
period matrix entries than those for the naive method.

%
%
%
%
%
%
%
\begin{table}[hbt]
\footnotesize
    \begin{center}
        \begin{tabular}{r|d|d|d|d|d|d}
& \multicolumn{3}{c|}{
$\Omega=\begin{pmatrix}\frac{-1+5i}2&\frac{i}6\\\frac{i}6&\frac{-1+7i}2\end{pmatrix}$}
& \multicolumn{3}{c}{
$\Omega=\begin{pmatrix}\frac{2+10i}7&\frac{1+2i}6\\\frac{1+2i}6&\frac{4}{10}+8i\end{pmatrix}$}\\
            \hline
bits
    & \multicolumn {1}{c|}{\magma}
    & \multicolumn {1}{c|}{\cmh-naive}
    & \multicolumn {1}{c|}{\cmh-Newton}
    & \multicolumn {1}{c|}{\magma}
    & \multicolumn {1}{c|}{\cmh-naive}
    & \multicolumn {1}{c}{\cmh-Newton} \\
\hline
$\approx 2^{11}$ &  0.46 &   0    &    0.02 &    0.03 &   0    &     0.02 \\
$\approx 2^{12}$ &  3.4  &   0.01 &    0.04 &    0.17 &   0.04 &     0.03 \\
$\approx 2^{13}$ &  26   &   0.07 &    0.08 &    1.1  &   0.20 &     0.09 \\
$\approx 2^{14}$ & 210   &   0.31 &    0.24 &    8.2  &   1.0  &     0.26 \\
$\approx 2^{15}$ &1700   &   1.3  &    0.69 &   60    &   5.2  &     0.75 \\
$\approx 2^{16}$ &       &   6.4  &    2.0  &  430    &   27   &     2.2  \\
$\approx 2^{17}$ &       &  32    &    5.7  & 3100    &  130   &     6.0  \\
$\approx 2^{18}$ &       & 160    &   16    &         &  720   &    16    \\
$\approx 2^{19}$ &       & 770    &   39    &         & 3100   &    40    \\
$\approx 2^{20}$ &       &3200    &   98    &         &        &    96    \\
$\approx 2^{21}$ &       &        &  240    &         &        &   230    \\
$\approx 2^{22}$ &       &        &  560    &         &        &   530    \\
$\approx 2^{23}$ &       &        & 1400    &         &        &  1300    \\
$\approx 2^{24}$ &       &        & 3200    &         &        &  3000    \\
$\approx 2^{25}$ &       &        & 7600    &         &        &  7100    \\
$\approx 2^{26}$ &       &        &16000    &         &        & 16000
        \end{tabular}
    \caption{\label{tab:theta-vs-magma}Computation of
    $\theta_0(\tau)$ (Intel i5-2500, 3.3GHz; \magma-2.19.4;
\cmh-1.0)}
    \end{center}
\end{table}

Notice that the running times for Newton lifts are consistent with the
theoretical complexity of $O (\mult (N) \log N)$.
The code in \cmh\ implements the
approach using finite differences for estimating the Jacobian matrix as
described in~\S\ref{ssec:theta-newton}, as well as an algorithm which
computes the exact Jacobian matrix along with the Borchardt mean as
given in \cite[Algorithme~16]{Dupont06}. Both converge equally well, but
the latter approach is computationally more expensive by roughly 45~\%,
accounted for by a larger number of multiplications.

\subsection{Breakdown of timings for small class polynomial examples}

\begin{table}[hbt]
\hfill
\begin{tabular}[t]{l|r}
\multicolumn{2}{c}{$K=\Q[X]/(X^4+144X^2+3500)$}\\
\multicolumn{2}{c}{$\Cf=\Norm_{\Phir}(\Cl_{\Kr})=\Z/2\Z\times \Z/30\Z$}\\
\hline
preparation & 0.2\\
\hline
base, $2\,000$ bits & 0.6\\
lift, $3\,984$ bits & 0.8\\
lift, $7\,944$ bits & 2.1\\
reconstruction attempt & 0.1\\
lift, $15\,846$ bits & 6.2\\
\\
\\
$H_1, \Hhat_2,\Hhat_3\in\C[X]$ & 0.1\\
$H_1, \Hhat_2,\Hhat_3\in K_0^r[X]$ & 3$\times$0.3\\
\hline
check & 0.8\\
\hline
Total (incl. I/O)& 12.4
\end{tabular}
\hfill
\begin{tabular}[t]{l|r}
\multicolumn{2}{c}{$K=\Q[X]/(X^4+134X^2+712)$}\\
\multicolumn{2}{c}{$\Cf=\Norm_{\Phir}(\Cl_{\Kr})=\Z/2\Z\times\Z/60\Z$}\\
\hline
preparation & 0.3\\
\hline
base, $2\,000$ bits& 1.1\\
lift, $3\,988$ bits& 1.6\\
lift, $7\,958$ bits& 4.4\\
reconstruction attempt & 0.1\\
lift, $15\,886$ bits& 13.1\\
reconstruction attempt & 0.2\\
lift, $31\,744$ bits& 38.7\\
$H_1, \Hhat_2,\Hhat_3\in\C[X]$ & 0.6\\
$H_1, \Hhat_2,\Hhat_3\in K_0^r[X]$ & 1.8 + 2$\times$1.4\\
\hline
check & 0.7\\
\hline
Total (incl. I/O) & 69.2
\end{tabular}
\hspace* {\fill}
\caption{\label{tab:simple}Timings in seconds for two examples
(on one Intel i5-2500, 3.3GHz)}
\end{table}

Table~\ref{tab:simple} illustrates our class polynomial computations on
relatively small examples.

Our code distinguishes orbits of the roots of the Igusa class polynomials
under complex conjugation. For instance, there are four real roots and
$58$ pairs of complex-conjugate roots in the second example, so that
altogether we need to carry out $62$ lifts of $\theta$-constants.
Instead of targeting a given precision based on
arguments as developed in~\cite{Streng09}, we simply carry out successive
lifting steps until the polynomial reconstruction succeeds. This explains
the time needed for failed reconstruction attempts in Table~\ref {tab:simple},
which could be avoided if we had a sharper bound on the required
precisions. It regularly occurs, even though this is not illustrated by the
examples here, that the reconstruction of the class polynomial
$H_1\in K_0^r[X]$ succeeds one lifting step before that of
$\Hhat_2,\Hhat_3\in K_0^r[X]$.
This can be explained by the relative size
of the invariants considered by Streng, see~\cite[Appendix 3]{Streng10}.

The timings indicated as ``preparation'' and ``check'' in
Table~\ref{tab:simple} correspond to the number theoretic calculations
performed in \parigp. The preparation time covers the enumeration of
$\Norm_{\Phir}(\Cl_{\Kr})\subseteq\Cf$, and the creation of the relevant set
of reduced period matrices. Checking means finding a Weil
number over a $128$-bit prime and generating a genus~$2$ curve whose Jacobian
has complex multiplication by the maximal order of~$K$.

\section {A large example}
\label{sec:large}

Our currently largest example is
$K = \Q [X] / (X^4 + 1357 X^2 + 3299)$, containing
$K_0 = \Q (\sqrt {1828253})$ of class number~$2$.
Its Shimura class group is
$\Cf = \Norm_{\Phir} (\Cl_{\Kr}) \simeq \Z / 2 \Z \times \Z / 2 \Z
\times \Z / 5004 \Z$ of size~$20016$.
On one core of an Intel Core i5-4570 clocked at $3.2$~GHz, the structure
of the class group is obtained with our \parigp\ script in roughly one second,
while the computation of the period matrices and their symbolic reduction
into the fundamental domain~$\Fc_2$ takes $388$~s.

The associated $\theta$-constants consist of $10008$ pairs of
complex-conjugate values. For the first ten Newton iterations up to a
precision of about $2\,000\,000$~bits, we used $640$ cores Intel Xeon X5675
at $3.07$~GHz; for the last two iterations, we switched to a machine with only
$160$ cores Intel Xeon E7-8837 at $2.67$~GHz, but with $640$~GB of main
memory.
Table~\ref{tab:lift-big} gives the timings (in seconds) for the Newton lifts of
one particular period matrix. The small value of $\delta$, estimated as
explained at the end of \S\ref {ssec:theta-newton},
shows that the effective precision indeed almost doubles in each step
as predicted by Theorem~\ref {th:newton}.

\begin{table}[hbt]
\begin {center}
\begin {tabular}{r|r|d}
precision & $\delta$ & \multicolumn {1}{c}{time} \\
\hline
     2\,000 &    --- &      0.03 \\
     3\,986 &     14 &      0.01 \\
     7\,970 &      2 &      0.1  \\
    15\,932 &      8 &      0.3  \\
    31\,862 &      2 &      1.0  \\
    63\,718 &      6 &      2.9  \\
   127\,434 &      2 &      8.2  \\
   254\,858 &     10 &     24    \\
   509\,714 &      2 &     61    \\
1\,019\,416 &     12 &    150    \\
2\,038\,832 &      0 &    360    \\
\hline
4\,077\,652 &     12 &    940   \\
8\,155\,302 &      2 & 2\,100
\end {tabular}
\caption{\label{tab:lift-big}Time for lifting steps for example 
with $\#\Cf=20\,016$.}
\end {center}
\end{table}

The lifting step accounts for a total of about $420$ CPU days, but thanks
to its easy parallelisation on $160$ to $640$ cores, it was finished in
less than $4$ days wall-clock time (including additional overhead for writing
intermediate results to disk).

The computation of the floating point polynomials $H_1$, $\Hhat_2$ and
$\Hhat_3$ was carried out at a precision of $7\,850\,071$ bits (the lowest
lifting precision reached for one of the period matrices).
After regrouping complex conjugates,
the first step consists of $5 \cdot 10008 / 2 = 25020$ multiplications of
monic polynomials of degree~$2$ with real coefficients, which can be arbitrarily
parallelised;
we used a machine with $24$ Intel Xeon E7540 cores
at $2.0$~GHz and $512$~GB of memory.
At degrees~$2\,048$ to~$8192$, the FFT multiplications required too much
memory to be executed in parallel on all
cores, so we reduced the number of simultaneous multiplications to
the maximum possible, as indicated in Table~\ref{tab:poly-big}.
In the last step, we needed to multiply a degree~$16\,384$ polynomial with
a degree~$3\,632$ polynomial.
The wall-clock time of this polynomial reconstruction step, counting
input-output time and the cost of resuming computations, was almost
exactly
$3$~days.

\begin{table}[hbt]
\begin {center}
\begin {tabular}{r|r|r|r}
input degree & $\#$ multiplications & wall-clock time (s) & \#threads\\
\hline
      2 & 25\,020 &     380 & 24 \\
      4 & 12\,510 &     560 & 24 \\
      8 &  6\,225 &     780 & 24 \\
     16 &  3\,125 &  1\,300 & 24 \\
     32 &  1\,565 &  1\,800 & 24 \\ %
     64 &     780 &  2\,700 & 24 \\ 
    128 &     390 &  3\,700 & 24 \\ 
    256 &     195 &  6\,300 & 24 \\ 
    512 &     100 & 11\,000 & 24 \\ 
 1\,024 &      50 & 13\,000 & 24 \\ 
\hline
 2\,048 &     25 & 25\,000 & 10\\ 
 4\,096 &     10 & 20\,000 & 10\\ 
\hline
 8\,192 &     5 & 71\,000 & 3 \\ 
\hline
16\,384 &     5 & 67\,000 & 5 \\ 
\end {tabular}
\caption{\label{tab:poly-big}Polynomial reconstruction timings for
    example
with $\#\Cf=20\,016$.}
\end {center}
\end{table}

Recognising one coefficient of the floating point polynomials
as an element of $\Kr_0$ took on average $980$~s per coefficient on one
Intel Xeon E5-2650 core at $2$~GHz.
The total CPU time for the $60\,045$ coefficients was thus about
$680$ CPU days; with up to $480$ cores working in parallel, this took
less than $2$~wall-clock days.

The uncompressed storage size of the three resulting polynomials in base~$10$
is about $90$~GB. The common denominator of the coefficients of $H_1$ has
$8\,884$ distinct prime factors, the largest one being $1\,506\,803\,839$.
It occurs to powers~$2$ in $H_1$ and~$4$ in $\Hhat_2$ and~$\Hhat_3$,
consistent with the fact that the power of $h_{10}$ in the denominator
of $j_2$ and $j_3$ is~$2$ instead of~$1$ for~$j_1$.

\subsection* {Acknowledgements}

The authors would like to acknowledge the work of Régis Dupont, whose
thesis has been extensively used as a basis from the outset of this work.
We are also grateful to Damien Robert and Marco Streng for fruitful
discussions and to David Gruenewald for his comments on an earlier version.

Computer experiments have used a variety of computing resources funded
from different projects. We thus acknowledge the support of the
Région
Lorraine CPER MISN TALC project;
the PlaFRIM experimental testbed, being developed under the Inria PlaFRIM
development action with support from LABRI and IMB and other entities:
Conseil Régional d'Aquitaine, FeDER, Université de Bordeaux and CNRS
(see \url {https://plafrim.bordeaux.inria.fr/});
the computing facilities MCIA (Mésocentre de Calcul Intensif Aquitain)
of the Université de Bordeaux and of the Université de Pau et des Pays
de l'Adour.

This research was partially funded by ERC Starting Grant ANTICS 278537
and by Agence Nationale de la Recherche grants ANR-09-BLAN-0020-01 and ANR-12-BS01-0010-01.

\bibliographystyle {plain}
\bibliography {cm2}

\end {document}